\documentclass[12pt]{article}
\usepackage{epsfig,graphicx}
\usepackage{fpcp03}

\catcode`\@=11
\renewcommand\footnoterule{%
  \kern-3\p@
  \hrule\@width.25\columnwidth
  \kern2.6\p@}
\catcode`@=12

\def\pageref#1{{?}}
\def\OMIT#1{{}}
\def\lsim{\mathrel{\!\mathpalette\vereq<}\!}

\def\vereq#1#2{\lower4pt\vbox{\baselineskip1.5pt \lineskip1.5pt
\ialign{$#1\hfill##\hfil$\crcr#2\crcr\sim\crcr}}}

\def\lqcd{\Lambda_{\rm QCD}}

\def\d{{\rm d}}

\def\GeV{{\rm GeV}}
\def\MeV{{\rm MeV}}

\newcommand{\nn}{\nonumber}
\newcommand{\beq}{\begin{equation}}
\newcommand{\eeq}{\end{equation}}
\newcommand{\beqa}{\begin{eqnarray}}
\newcommand{\eeqa}{\end{eqnarray}}

\newcommand{\Vub}{\ensuremath{V_{ub}}}
\newcommand{\Vcb}{\ensuremath{V_{cb}}}

\newcommand{\Bbar}{\,\overline{\!B}}
\newcommand{\Dbar}{\,\overline{\!D}}
\newcommand{\Kbar}{\,\overline{\!K}}
\def\B0bar{\Bbar{}^0}
\def\D0bar{\Dbar{}^0}
\def\K0bar{\Kbar{}^0}

\def\FD{{\cal F}}
\def\FDs{\FD_*}
\def\FDt{\FD_{(*)}}

\begin{document}


\Title{\boldmath $|V_{cb}|$ and $|V_{ub}|$: Theoretical Developments%
\,\footnotemark}
\bigskip

{\footnotetext{Invited talk at Flavor Physics \& CP Violation
(FPCP 2003), June 2003, Ecole Polytechnique, Paris, France.}}
 

%
\label{LigetiStart}

%
\author{Zoltan Ligeti\index{Ligeti, Z.}}

%
\address{Ernest Orlando Lawrence Berkeley National Laboratory\\
  University of California, Berkeley, CA 94720, USA}

\makeauthor

\abstracts{
The determinations of $|V_{cb}|$ and $|V_{ub}|$ from semileptonic $B$ decays
are reviewed with emphasis on recent developments and theoretical
uncertainties.  Future prospects and limitations are also discussed.
\hfill {\footnotesize [hep-ph/0309219, LBNL--53298]}}

\section{Introduction}

Since $\sin2\beta$, the $CP$ asymmetry in $B\to \psi K_S$ and related
modes~\cite{babelle}, is consistent with $\epsilon_K$, $|\Vub|$, and $B_{d,s}$
mixing, searching for new flavor physics will require a combination of
``redundant" and precise measurements that relate in the Standard Model (SM) to
the same CKM elements.  The determination of $|\Vcb|$ is important because the
uncertainties in $\epsilon_K$ and in $K\to \pi\nu\bar\nu$ are proportional to
$|\Vcb|^4$, while the uncertainty of $|V_{ub}|$ dominates the error of a side
of the unitarity triangle.  Some of the best strategies to look for new physics
include comparing angles and sides of the unitarity triangle, and results from
tree and loop processes, and semileptonic decays are crucial for this. 
Processes mediated by flavor-changing neutral currents, such as $b\to
q\,\gamma$, $b\to q\, \ell^+\ell^-$, and $b\to q\, \nu\bar\nu$ ($q=s,d$) are
sensitive probes of the SM, and the theoretical tools to analyze these are the
same as for $|V_{xb}|$.  It is the accuracy of the theory that ultimately
limits the sensitivity to new physics~\cite{SSI}.

\begin{figure}[bt]
\centerline{\includegraphics*[width=.5\textwidth]{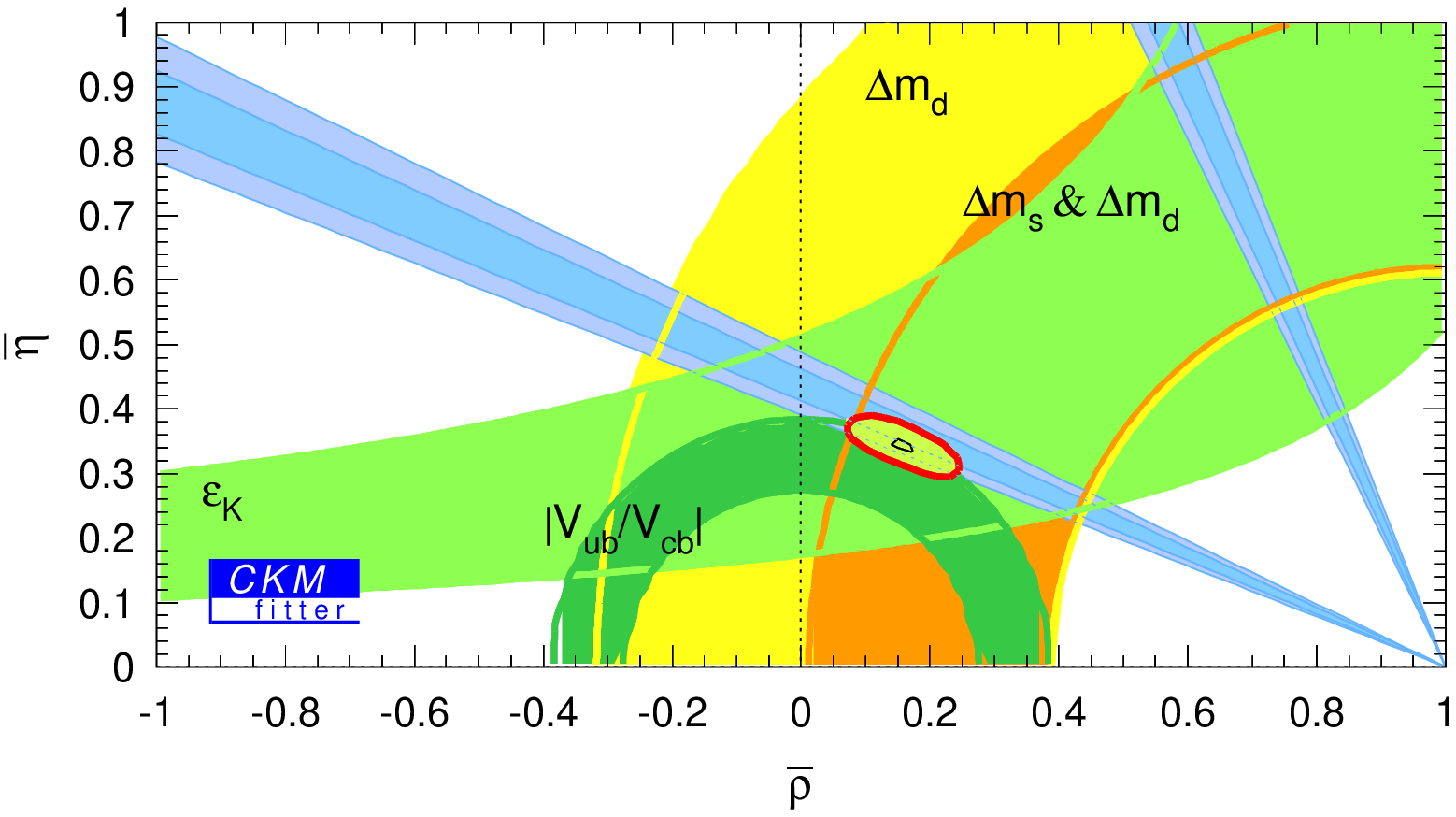}
\includegraphics*[width=.5\textwidth]{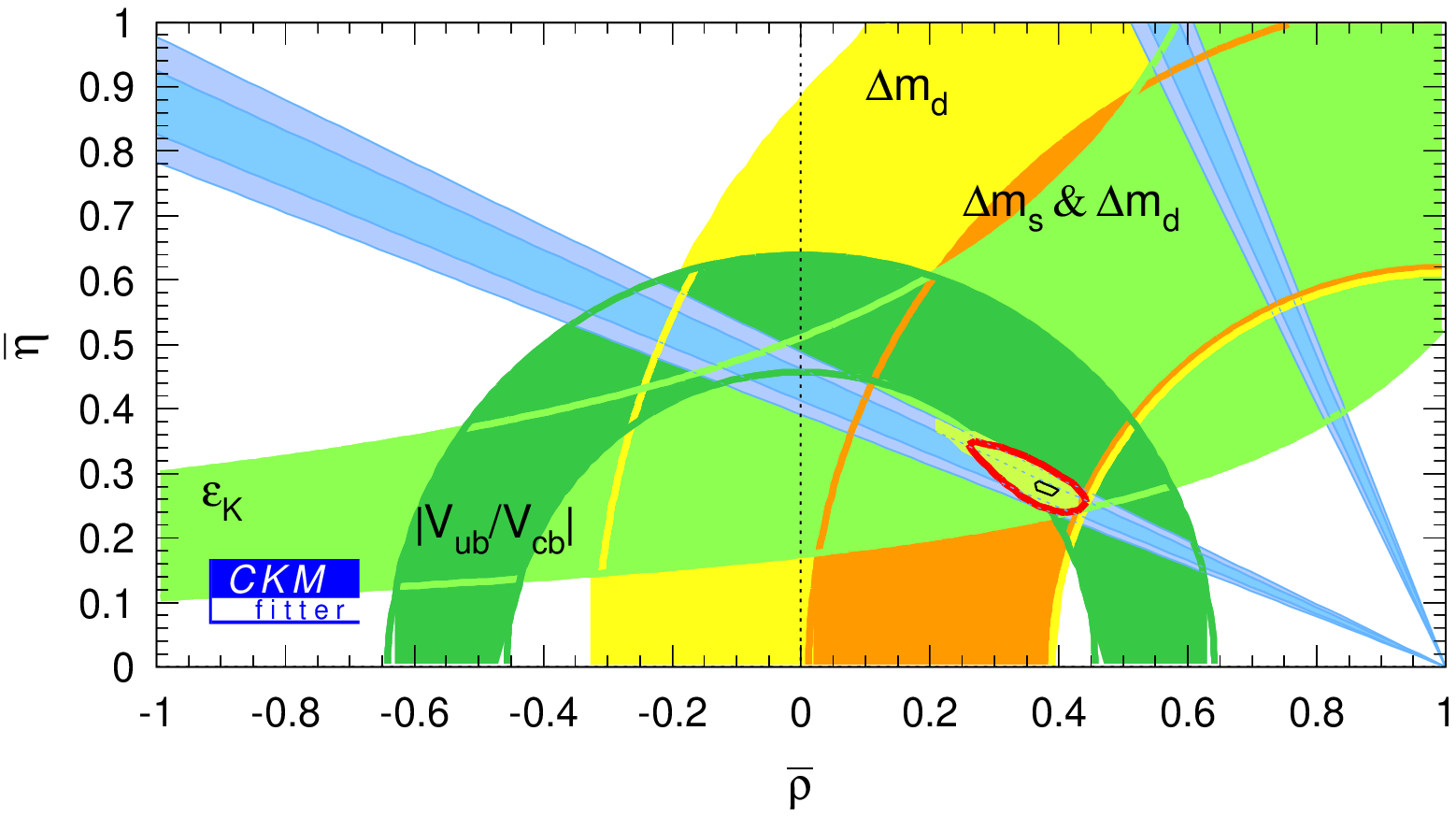}}
\caption{CKM fits with hypothetical values, $|V_{ub}| = (3.0 \pm 0.15 \pm 0.15)
\times 10^{-3}$ (left) and $(5.0 \pm 0.25 \pm 0.25) \times 10^{-3}$ (right),
and $\sin2\beta =$ its present world average with half the
error~\cite{ckmfitter}.}
\label{fig:intro}
\end{figure}

To illustrate where the future might take us, Fig.~\ref{fig:intro} shows CKM
fits assuming that $\sin2\beta$ equals its present central value with half the
error, and that $|\Vub|$ is about $1.5\sigma$ lower [higher] than its central
value with 5\% experimental and theoretical errors: $|V_{ub}| = (3.0 \pm 0.15
\pm 0.15) \times 10^{-3}$ [$(5.0 \pm 0.25 \pm 0.25) \times 10^{-3}$].  These
fits are motivated by the fact that recent exclusive [inclusive] measurements
of $|V_{ub}|$ appear to be on the low [high] side~\cite{thorndike}.  The
resulting central values of the angle $\gamma$ differ by $25^\circ$, and the SM
value of $\Delta m_s$ is near the minimum [maximum] of its presently allowed
range.  Clearly, an accuracy of $\sigma(|V_{ub}|) \sim 5\%$ is very desirable.

To believe at some point in the future that a discrepancy between measurements
is due to new physics, model independent predictions are crucial.  Results that
depend on modeling nonperturbative strong interaction effects cannot prove that
there is new flavor physics beyond the SM.  Model independent predictions are
those where the theoretical uncertainties are suppressed by powers of small
parameters, typically $\lqcd/m_b$, $m_s/\Lambda_{\chi SB}$, $\alpha_s(m_b)$,
etc.  Still, in most cases, there are uncertainties at some order, which cannot
be estimated model independently.  If the goal is to test the Standard Model,
one must assign sizable uncertainties to such ``small corrections" not known
from first principles.

Over the last decade most of the theoretical progress in understanding $B$
decays utilized that $m_b$ is much larger than $\lqcd$.  However, depending on
the process under consideration, the relevant hadronic scale may or may not be
numerically much smaller than $m_b$ (and, especially, $m_c$).  For example,
$f_\pi$, $m_\rho$, and $m_K^2/m_s$ are all of order $\lqcd$, but their
numerical values span an order of magnitude.  In most cases experimental
guidance is needed to decide how well the theory works for various processes.

\section{\boldmath $B\to X_c \ell\bar\nu$}

\subsection{\boldmath $|V_{cb}|$ exclusive: Heavy Quark Symmetry}

In mesons composed of a heavy quark and a light antiquark (plus gluons and
$q\bar q$ pairs), the energy scale of strong interactions is small compared to
the heavy quark mass.  The heavy quark acts as a static point-like color source
with fixed four-velocity, which cannot be altered by the soft gluons
responsible for confinement.  Thus, the configuration of the light degrees of
freedom (``brown muck") become insensitive to the spin and flavor (mass) of the
heavy quark. 

Heavy quark symmetry (HQS)~\cite{HQS} is especially predictive for $B\to
D^{(*)}$ semileptonic decays.  When the weak current changes suddenly (on a
time scale $\ll \lqcd^{-1}$) the flavor $b\to c$, the momentum $\vec p_b \to
\vec p_c$, and possibly the spin, $\vec s_b\to \vec s_c$, the brown muck only
feels that the four-velocity of the static color source in the center of the
meson changed, $v_b \to v_c$.  Therefore, the form factors that describe the
wave function overlap between the initial and final mesons become independent
of the Dirac structure of weak current, and can only depend on a scalar
quantity, $w \equiv v_b \cdot v_c$.  Thus all form factors are related to a
single Isgur-Wise function, $\xi(v_b\cdot v_c)$, which contains all the low
energy nonperturbative hadronic physics relevant for these decays.  Moreover,
$\xi(1)=1$, because at ``zero recoil", $w=1$, where the $c$ quark is at rest in
the $b$ quark's rest frame, the configuration of the brown muck does not change
at all.

The determination of $|V_{cb}|$ from exclusive $B\to D^{(*)} \ell \bar\nu$
decay uses an extrapolation of the measured rate to zero recoil, $w=1$. The
rates can be schematically written as
\begin{equation}\label{rates}
{\d\Gamma(B\to D^{(*)} \ell\bar\nu)\over \d w} = (\mbox{known factors})\,
  |V_{cb}|^2\, \cases{ (w^2-1)^{1/2}\, \FDs^2(w)\,,\quad & for $B\to D^*$, \cr
  (w^2-1)^{3/2}\, \FD^2(w)\,, & for $B\to D$\,. \cr}
\end{equation}
In the heavy quark limit $\FD(w) = \FDs(w) = \xi(w)$, and in particular
$\FDt(1) = 1$, allowing for a model independent determination of $|V_{cb}|$.  
The corrections to the $m_Q \to\infty$ limit ($Q=b,c$) can be organized in a
simultaneous expansion in $\alpha_s(m_Q)$ and $\lqcd/m_Q$, of the form
\beqa\label{F1}
\FDs(1) &=& 1_{\mbox{\footnotesize (Isgur-Wise)}} + c_A(\alpha_s) 
  + {0_{\mbox{\footnotesize (Luke)}}\over m_Q} 
  + {(\mbox{lattice or models})\over m_Q^2} + \ldots \,, \nn\\
\FD(1) &=& 1_{\mbox{\footnotesize (Isgur-Wise)}} + c_V(\alpha_s) 
  + {(\mbox{lattice or models})\over m_Q} + \ldots \,.
\eeqa
The perturbative corrections $c_A = -0.04$ and $c_V = 0.02$ are known to
order $\alpha_s^2$~\cite{Czar}, and higher order terms should be below
the $1\%$ level.  The order $\lqcd/m_Q$ correction to $\FDs(1)$ vanishes due to
Luke's theorem~\cite{Luke}.  The terms indicated by $(\mbox{lattice or
models})$ are only known using phenomenological models or quenched lattice QCD
at present.  This is why the determination of $|V_{cb}|$ is more reliable from
$B\to D^* \ell\bar\nu$ than from $B\to D\ell\bar\nu$, although both QCD sum
rules~\cite{LNN} and quenched lattice QCD~\cite{latticeD} suggest that the
$\lqcd/m_{b,c}$ correction to $\FD(1)$ is small (giving $\FD(1) = 1.02 \pm
0.08$ and $1.06 \pm 0.02$, respectively).  The rate near $w=1$ is larger in
$B\to D^*\ell\bar\nu$ than in $B\to D\ell\bar\nu$, because of the $w^2-1$
helicity suppression of the latter, yielding~\cite{hfag}
\beq\label{zerorec}
|V_{cb}|\, \FDs(1) = (36.7 \pm 0.8) \times 10^{-3} \,,\qquad
|V_{cb}|\, \FD(1) = (42.1 \pm 3.7) \times 10^{-3} \,.
\eeq
Using $\FDs(1) = 0.91 \pm 0.04$ (an estimate unchanged for many
years~\cite{babook} and supported by a recent quenched lattice
calculation~\cite{latticeDs}), yields $|V_{cb}| = (40.2 \pm 0.9_{\rm exp} \pm
1.8_{\rm th}) \times 10^{-3}$.  The $B\to D\ell\bar\nu$ data is consistent, but
to make a real test (and to further reduce the theoretical error of $|\Vcb|$), 
unquenched lattice calculations of $\FDt(1)$ are needed.

Another important theoretical input is the shape of $\FDt(w)$ used for fitting
the data.  Expanding about zero recoil, one writes $\FDt(w) = \FDt(1)\, [1 -
\rho_{(*)}^2 (w-1) + c_{(*)} (w-1)^2 + \ldots]$.  Knowing the slope,
$\rho_{(*)}^2$, is important because it has a large correlation with the
extracted value of $|V_{cb}|\, {\cal F}_*(1)$.  Analyticity imposes stringent
constraints between $\rho_{(*)}^2$ and the curvature, $c_{(*)}$~\cite{BGL},
which is used in the fits to obtain Eq.~(\ref{zerorec}).  The $B\to
D\ell\bar\nu$ measurement is also important, because  HQS constrains the
differences $\rho_*^2-\rho^2$ and $c_*-c$~\cite{BGZL}, and computing $\FD(1)$
on the lattice is not harder than $\FDs(1)$.  Sum rules have also been used to
constrain $\FDt(1)$ and the slope parameter~\cite{sumrules}, and very recently
a new set of recursive bounds on all derivatives of the Isgur-Wise function at
zero recoil were obtained~\cite{orsay}
\beq
(-1)^n\, \xi^{(n)}(1) \geq {2n+1\over 4}\, \Big[(-1)^{n-1}\, \xi^{(n-1)}(1)
\Big] \quad \Rightarrow \quad
(-1)^n\, \xi^{(n)}(1) \geq {(2n+1)!!\over 2^{2n}}\,.
\eeq
An important ingredient in the sum rules are the excited states' contributions,
so their precise understanding will improve the determination of $|V_{cb}|$
from both exclusive and inclusive decays.  How to best use all the information
on the shapes (of both the $B\to D^*\ell\bar\nu$ form factors and the
$w$-spectra) still appears a somewhat open question where progress could be
made.

\subsection{\boldmath $|V_{cb}|$ inclusive: OPE and HQET parameters}

In the large $m_b$ limit, there is a simple argument based on a separation of
scales that the inclusive rate may be modeled by the decay of a free $b$
quark.  The $b$ quark decay mediated by the weak interaction takes place on a
time scale that is much shorter than the time it takes the quarks in the final
state to hadronize.  Once the $b$ quark has decayed, the probability that the
decay products will hadronize somehow is unity, and we need not know the
(uncalculable) probabilities of hadronization to specific final states.  

The above argument can be made precise using an operator product expansion
(OPE).  When the energy release to the final hadronic state is large, the
forward scattering amplitude (whose imaginary part gives the decay rate) can be
expanded in local operators,
\beq\label{opesketch}
\raisebox{-44pt}{\includegraphics*[width=.34\textwidth]{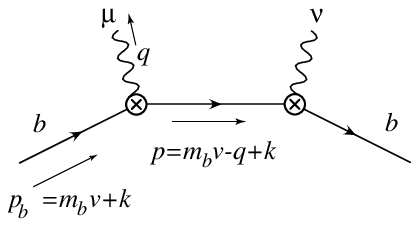}} = 
  \raisebox{-22pt}{\includegraphics*[width=.12\textwidth]{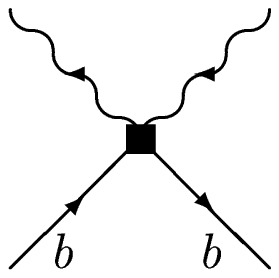}}
 + {0\over m_b} \raisebox{-22pt}{\includegraphics*[width=.12\textwidth]{ope2}}
 + {1\over m_b^2} \raisebox{-22pt}{\includegraphics*[width=.12\textwidth]{ope2}}
 + \ldots \,.
\eeq
The OPE allows the model independent computation of inclusive semileptonic $B$
decay rates in a series in $\lqcd/m_b$ and $\alpha_s(m_b)$~\cite{OPE},
schematically as
\beq\label{incl}
\d\Gamma = \pmatrix{b{\rm ~quark} \cr {\rm decay}\cr} \times 
\bigg[ 1 + \frac0{m_b} + \frac{f(\lambda_1,\lambda_2)}{m_B^2} + \ldots
  + \alpha_s(\ldots) + \alpha_s^2(\ldots) + \ldots \bigg] \,.
\eeq
It proves that semileptonic rates in the $m_b\to\infty$ limit are given by $b$
quark decay, and the leading nonperturbative corrections suppressed by
$\lqcd^2/m_b^2$ can be parameterized by two HQET matrix elements, $\lambda_1$
and $\lambda_2$.  Most quantities of interest have been computed including the
order $\lqcd^3/m_b^3$ nonperturbative corrections (which are parameterized by
six more hadronic matrix elements, $\rho_{1,2}$ and $\tau_{1,\ldots 4}$), while
the perturbation series at leading order in $\lqcd/m_b$ are known including the
$\alpha_s$ and $\alpha_s^2\beta_0$ terms ($\beta_0 = 11 - 2n_f/3$ is the first
coefficient of the QCD $\beta$-function, and this term often dominates at order
$\alpha_s^2$).

For fully inclusive quantities, such as the $B\to X_c \ell\bar\nu$ rate, the
OPE calculation should be under good control.  The theoretical uncertainties in
this case come from the error in a short distance $b$ quark mass (whatever way
it is defined), the perturbation series, and the nonperturbative corrections. 
There has been a lot of recent progress in determining a correlated range of
$|\Vcb|$, $m_b$, $\lambda_1$ and the HQET matrix elements at order
$\lqcd^3/m_b^3$ from measurements of shape variables.  The idea is to compare
the OPE predictions with data for shapes of decay distributions (spectral
moments) that are independent of CKM elements.  Since different spectra have
different dependence on $m_b$, $\lambda_1$, etc., a simultaneous fit to many
moments and to the semileptonic width allows both the determination of the
hadronic parameters and $|V_{cb}|$, and tests the validity of the whole
approach.  The observables which received the most attention are the charged
lepton energy~\cite{gremmetal,GK} and hadronic invariant mass~\cite{FLSmass,GK}
spectra in $B\to X_c\ell\bar\nu$, and the photon energy spectrum in $B\to
X_s\gamma$~\cite{kl}.  Their measurements show improving consistency, and were
discussed  elsewhere at this conference~\cite{artusocalvi}.

Two recent global fits used somewhat different, but in principle equivalent,
approaches.  In Ref.~\cite{shape} both the $B^*-B$ and $D^*-D$ mass differences
are used to constrain linear combinations of $\lambda_2$ and some of the
$\lqcd^3/m_b^3$ matrix elements, and the fit contained seven unknowns:
$|V_{cb}|,\ m_b,\ \lambda_1,\ \rho_1,\ \tau_1-3\tau_4,\ \tau_2+\tau_4,\
\tau_3+3\tau_4$.  In Ref.~\cite{delphifit} no expansion in $m_c$ is performed,
and in this case the seven free parameters are: $|V_{cb}|,\ m_b,\ m_c,\
\lambda_1,\ \lambda_2,\ \rho_1,\ \rho_2$ (note that Ref.~\cite{delphifit}
fitted fewer parameters).  The difference between the two approaches is of
order $\lqcd^4/(m_c^2 m_b^2)$.  The results are $|V_{cb}| = (40.8 \pm 0.9)
\times 10^{-3}$~\cite{shape} and $|V_{cb}| = (41.1 \pm 1.1) \times
10^{-3}$~\cite{delphifit}.

Comparing these shape variables is also the most promising approach to
constrain experimentally the accuracy of OPE, including possible quark-hadron
duality violation.  Quark-hadron duality~\cite{PQW} is the notion that averaged
over exclusive channels, hadronic quantities can be computed at the parton
level.  This is implicitly assumed in the OPE.  Duality violations are believed
to be small for fully inclusive semileptonic $B$ decay rates, however, it is
hard to quantify how small~\cite{NIdual}.  One can further test the theory by
weighing the lepton spectrum with suitably chosen fractional powers of $E_\ell$
to reduce the nonperturbative corrections.  As shown in
Table~\ref{tab:btmoments}, predictions and data for such ``Bauer-Trott
moments"~\cite{BT} are in excellent agreement.

\begin{table}[htbp]
\centerline{\footnotesize
\begin{tabular}{|cccccc|}
\hline
$R_{3a}$ & $R_{3b}$ & $R_{4a}$ & $R_{4b}$ & $D_3$ & $D_4$ \\ \hline
$0.302\pm0.003$ & $2.261\pm0.013$ & $2.127 \pm0.013$ & $0.684\pm0.002$ 
  & $0.520\pm0.002$ & $0.604\pm0.002$ \\ 
$0.3016\pm0.0007$ & $2.2621\pm0.0031$ & $2.1285\pm0.0030$ 
  & $0.6833\pm0.0008$ & $0.5193\pm0.0008$ & $0.6036\pm0.0006$ \\ \hline
\end{tabular}
}
\caption{Predictions~\cite{shape} (above) and data~\cite{cleoBT} (below) for
Bauer-Trott moments.}
\label{tab:btmoments}
\end{table}

In the future it will be important to determine the $B\to D^{(*)}\ell\bar\nu$
branching ratios with higher precision, to model independently map out the
higher mass charm states in semileptonic $B$ decay, and to measure the $B\to
X_s\gamma$ spectrum to as low photon energies as possible.  Completing the full
two-loop calculation of spectra would also be useful. If these measurements are
consistent as they get very precise, the theoretical limitation appears to be
around $\sigma (|V_{cb}|) \approx 3.5 \times 10^{-4}$ and $\sigma (m_b^{1S})
\approx 35\,\MeV$~\cite{shape}.

\section{\boldmath $B\to X_u \ell\bar\nu$}

\subsection{\boldmath $|V_{ub}|$ exclusive: few comments}

In $B$ decays to light mesons there is a much more limited use of heavy quark
symmetry than in $B\to D^{(*)}$, since it does not apply for the final state. 
One can still derive relations between the $B\to \rho\ell\bar\nu$,
$K^*\ell^+\ell^-$, and $K^*\gamma$ form factors in the large $q^2$
region~\cite{IsWi}.  In the small $q^2$ region when the energy of the light
hadron is large, Soft-Collinear Effective Theory (SCET)~\cite{SCET} can be used
to prove a factorization theorem for form factors~\cite{SCETsl}.  The
nonfactorizable part satisfies form factor relations~\cite{charles}, while the
factorizable part does not.  These two terms are of the same order in
$\lqcd/m_b$.  How they compare numerically, or in the $m_b\to\infty$ limit when
effects of order $\alpha_s(m_b)$ and $\alpha_s(\sqrt{m_b\lqcd})$ are fully
accounted for is an open question. 

To determine $|V_{ub}|$ with sub-10\% error, model independent determination of
the form factors is needed.  This can be achieved in future unquenched lattice
QCD calculations, as discussed elsewhere at this conference~\cite{damir}. 
Another possibility is to combine heavy quark and chiral symmetries to form
``Grinstein-type double ratios"~\cite{Gtdr}, whose deviation from unity
is suppressed in both symmetry limits.  For example,
\beq
{f_B\over f_{B_s}} \times {f_{D_s}\over f_D}
= 1 + {\cal O}\bigg( {m_s\over m_c}-{m_s\over m_b}\,,\;
  {m_s\over 1\,{\rm GeV}}\, {\alpha_s(m_c)-\alpha_s(m_b)\over\pi} \bigg) \,,
\eeq
and lattice calculations indicate that the deviation from unity is at the few 
percent level.  If three quantities on the left-hand side are measured, the
forth is determined with small uncertainty.  Similar double ratios can be
constructed for the semileptonic decay form factors~\cite{lw}
\beq
{f^{(B\to\rho\ell\bar\nu)} \over f^{(B\to K^*\ell^+\ell^-)}} \times 
{f^{(D\to K^*\ell\bar\nu)} \over f^{(D\to\rho\ell\bar\nu)}}\,,
\eeq
or for appropriately weighted $q^2$-spectra in these decays, and may be
experimentally  accessible soon.  Recently the leading power corrections to the
HQS relations between the $B$ and $D$ decay form factors were
analyzed~\cite{BGDP}.  Replacing $K^*\ell^+\ell^-$ by $K^*\nu\bar\nu$ that may
be accessible at a super-B-factory would make the deviation of this double
ratio from unity very small.  With data from hadronic- and super-$B$-factories
the double ratio~\cite{ringberg}
\beq
{{\cal B}(B\to\ell\bar\nu) \over {\cal B}(B_s\to \ell^+\ell^-)} \times 
{{\cal B}(D_s\to \ell\bar\nu) \over {\cal B}(D\to\ell\bar\nu)}\,,
\eeq
could give a determination of $|V_{ub}|$ with theoretical errors at the few
percent level.

\subsection{\boldmath $|V_{ub}|$ inclusive: cuts on $B\to X_u\ell\bar\nu$
spectra}

If it were not for the huge $B\to X_c\ell\bar\nu$ background which is
$\sim$\,50 times larger than the $B\to X_u\ell\bar\nu$ signal, measuring
$|V_{ub}|$ would be as ``easy" as $|V_{cb}|$.  The fully inclusive $B\to X_u
\ell\bar\nu$ rate can be calculated in the OPE with small
uncertainty~\cite{upsexp},
\begin{equation}\label{Vub}
|V_{ub}| = (3.04 \pm 0.06_{(\rm pert)} \pm 0.08_{(m_b)}) \times 10^{-3}\,
  \bigg( {{\cal B}(B\to X_u \ell\bar\nu)\over 0.001}
  {1.6\,{\rm ps}\over\tau_B} \bigg)^{1/2} ,
\end{equation}
where the first error is from the perturbation series and  the second is from
the $b$ quark mass, $m_b^{1S} = 4.73 \pm 0.05\,$GeV.  If this rate is measured
without significant cuts on the phase space, then $|V_{ub}|$ can be determined
with less than $5\%$ theoretical error.  

\begin{figure}[t]
\centerline{\includegraphics*[width=.95\textwidth]{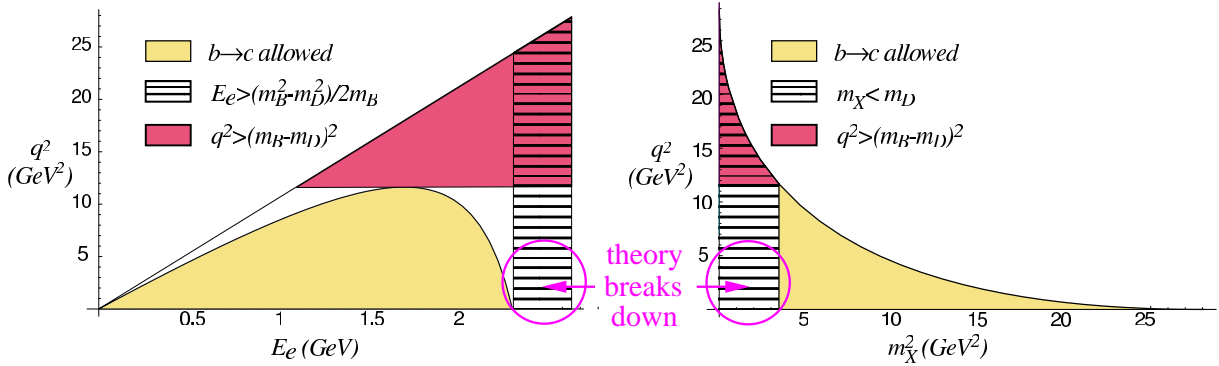}}
\caption{Dalitz plots for $B\to X\ell\bar\nu$ in terms of $E_\ell$ and
$q^2$ (left), and $m_X^2$ and $q^2$ (right).}
\label{fig:dalitz}
\end{figure}

The behavior of the OPE can become significantly worse if kinematic cuts are
imposed to distinguish the $b\to u$ signal from the $b\to c$ background.  All
such proposed cuts imply (directly or indirectly) $m_X < m_D$.  Even if many
different resonances can be produced in the final state, and therefore the
inclusive description is expected to be appropriate, such cuts may still
distroy the convergence of the OPE.  One may think of the OPE as an expansion
of the diagram on the left-hand side of Eq.~(\ref{opesketch}) in powers of $k$
(which is of order $\lqcd$), the residual momentum of the $b$ quark in the $B$
meson,
\beq
{1\over (m_b v + k -q)^2} 
  = {1\over (m_b v -q)^2 + 2k\cdot (m_b v-q) + k^2} \,.
\eeq
In the $m_b \gg \lqcd$ limit this expansion converges in most of the phase
space.  If cuts are applied to the final state phase space, the expansion in
$k$ only converges if the three terms of different orders in $k$ on the
right-hand side exhibit a hierarchy.  For $m_X \ll m_B$ this implies that the
range of hadronic final states that are allowed to contribute should satisfy
\beq\label{converge}
m_X^2 \gg E_X \lqcd \gg \lqcd^2 \,.
\eeq
Thus, depending on whether the allowed invariant mass and energy of the
hadronic final state (in the $B$ rest frame) satisfies Eq.~(\ref{converge}), 
there are three qualitatively different regions:
\begin{itemize} \vspace*{-7pt}\itemsep -4pt

\item[(i)] $m_X^2 \gg E_X \lqcd \gg \lqcd^2$: the OPE converges, and the first
few terms are expected to give reliable result.  (This is the case for the 
$B\to X_c\ell\bar\nu$ rate relevant for measuring $|V_{cb}|$.)

\item[(ii)] $m_X^2 \sim E_X \lqcd \gg \lqcd^2$: an infinite set of equally
important terms in the OPE must be resummed.  The OPE becomes a twist expansion
and nonperturbative input is needed.

\item[(iii)] $m_X \sim \lqcd$: the final state is dominated by resonances, and
it is not known how to compute inclusive quantities reliably.

\end{itemize}\vspace*{-7pt}
The charm background can be removed by several different kinematic cuts:
\begin{enumerate} \vspace*{-7pt}\itemsep -4pt

\item $E_\ell > (m_B^2-m_D^2) / (2m_B)$: the endpoint region of the charged
lepton energy spectrum;

\item $m_X < m_D$: the small hadronic invariant mass
region~\cite{mass,FLW,BDU,llrhadron};

\item $E_X < m_D$: the small hadronic energy region~\cite{energy};

\item $q^2 \equiv (p_\ell + p_\nu)^2 > (m_B - m_D)^2$: the large dilepton
invariant mass region~\cite{BLL1}.

\end{enumerate}\vspace*{-7pt}
These contain roughly $10\%$, $80\%$, $30\%$, and $20\%$ of the rate,
respectively.  Measuring any other variable than $E_\ell$ requires the
reconstruction of the neutrino momentum, which is challenging experimentally. 
Combinations of cuts have also been proposed, $q^2$ with $m_X$~\cite{BLL2},
$q^2$ with $E_\ell$~\cite{KoMe}, or $m_X$ with $E_X$~\cite{ugo}.  
These regions of the Dalitz plot are shown in Fig.~\ref{fig:dalitz}.

The problem is that both phase space regions 1.\ and 2.\ belong to the regime
(ii), because these cuts allow $m_X$ up to ${\cal O}(m_D)$ and $E_X$ up to 
${\cal O}(m_B)$, and numerically $\lqcd\, m_B \sim m_D^2$.  The region $m_X <
m_D$ is better than $E_\ell > (m_B^2-m_D^2) / (2m_B)$ inasmuch as the expected
rate is larger, and the inclusive description is expected to hold better.  But
nonperturbative input is needed formally at the ${\cal O}(1)$ level in both
cases, which is why the model dependence of the $|V_{ub}|$ measurement from the
$m_X$ spectrum increases rapidly as the $m_X$ cut is lowered below
$m_D$~\cite{FLW}.  

\begin{figure}[t]
\centerline{\includegraphics*[height=3.4cm]{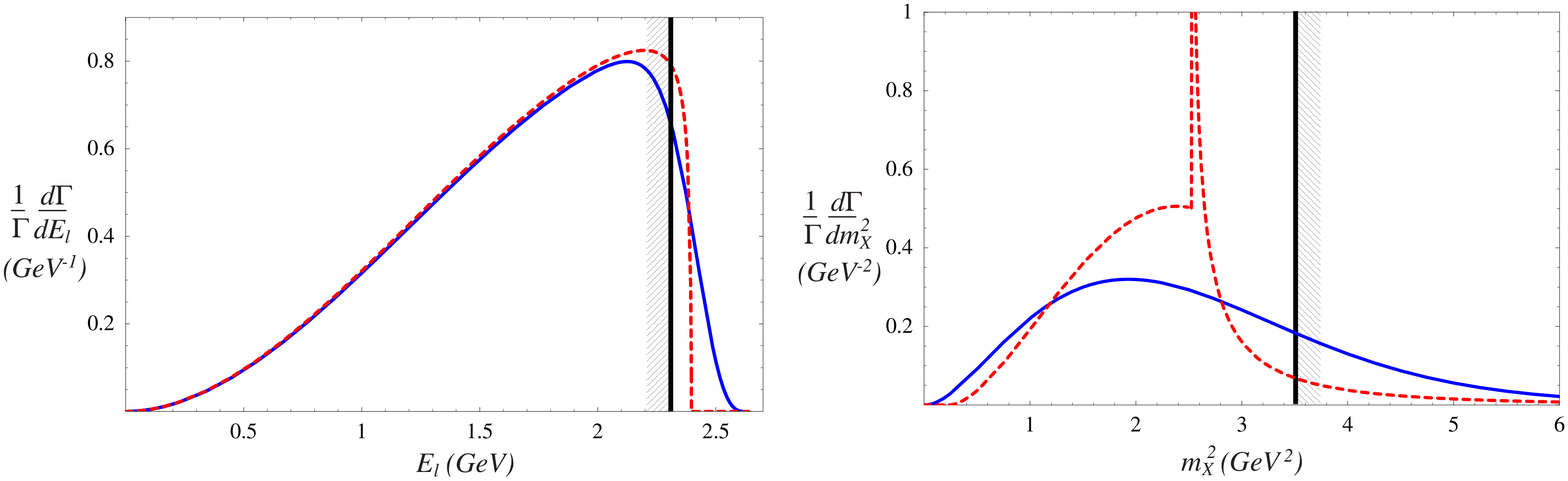}~
\includegraphics*[height=3.38cm]{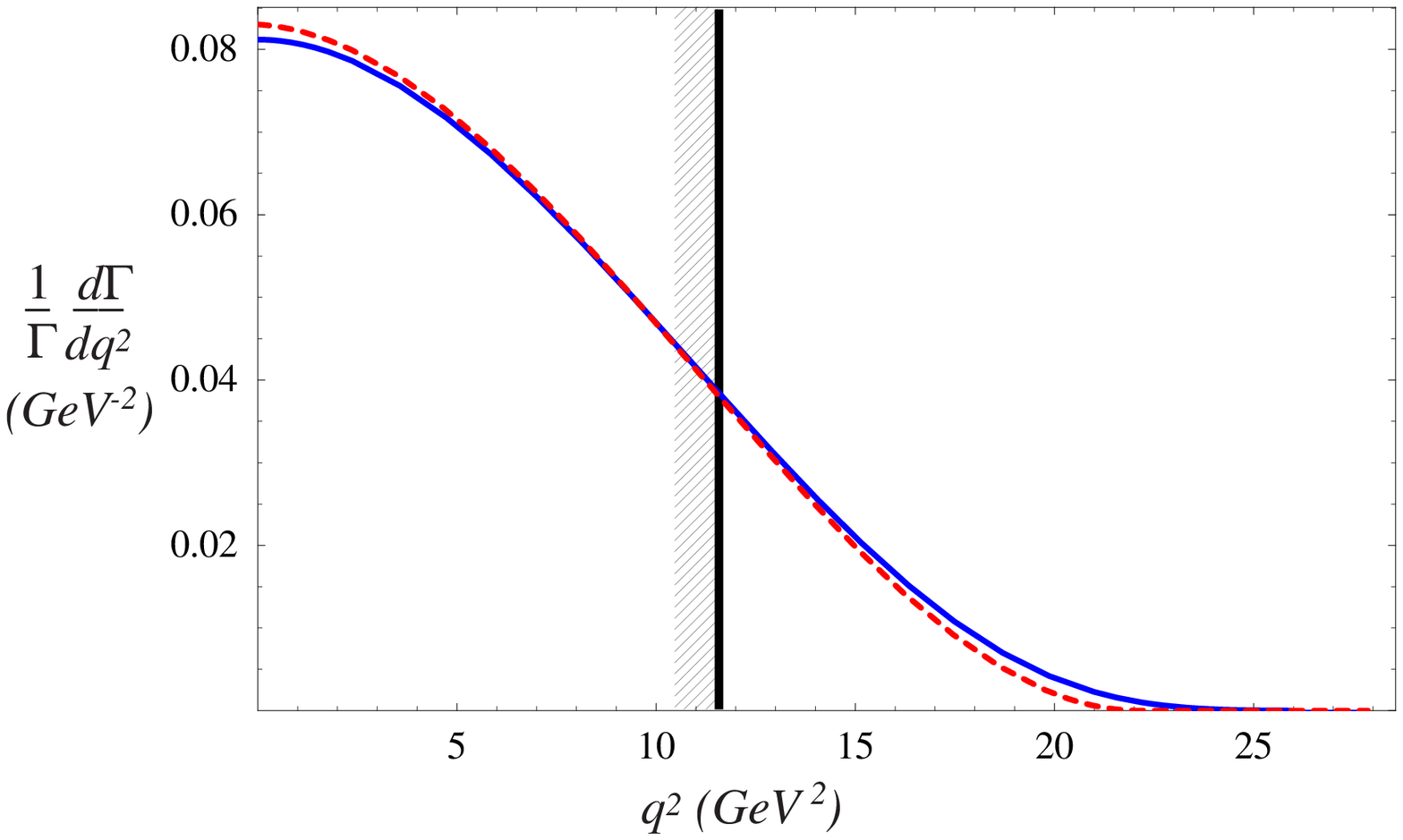}}
\caption{$E_\ell$ (left), $m_X^2$ (center), and $q^2$ (right) spectra. 
Dashed curves show $b$ quark decay to order $\alpha_s$, solid curves
include a Fermi motion model, shading shows the $b\to c$ kinematic limit.}
\label{fig:smear}
\end{figure}

The spectrum in the large $E_\ell$ and small $m_X$ regions are determined by
the $b$ quark light-cone distribution function that describes the Fermi motion
of the $b$ quark inside the $B$ meson (sometimes called the shape function). 
Its effect on the spectra are illustrated in Fig.~\ref{fig:smear}, where we
also show the $q^2$ spectrum unaffected by it.  This nonperturbative function
is universal at leading order in $\lqcd/m_b$, and is related to the $B\to
X_s\gamma$ photon spectrum~\cite{structure}.  These relations have been
extended to the resummed next-to-leading order corrections~\cite{extractshape},
and to include effects of operators other than $O_7$ contributing to $B\to
X_s\gamma$~\cite{notO7}.  Weighted integrals of the $B\to X_s\gamma$ photon
spectrum are related to the $B\to X_u\ell\bar\nu$ rate in the large $E_\ell$ or
small $m_X$ regions.  Recently CLEO used the $B\to X_s\gamma$ photon spectrum
as an input to determine $|V_{ub}| = (4.08 \pm 0.63)\times
10^{-3}$~\cite{cleoVub} from the lepton endpoint region.

The dominant theoretical uncertainty in this determination of $|V_{ub}|$ is
from subleading twist contributions suppressed by $\lqcd/m_b$, which are not
related to $B\to X_s\gamma$~\cite{subltwistus}.  The $B\to X_u \ell\bar\nu$
lepton spectrum, including dimension-5 operators and neglecting perturbative
corrections is~\cite{OPE}
\beqa\label{slspec}
{\d \Gamma\over \d y} &=& {G_F^2\, m_b^5\, |V_{ub}|^2\, \over 192\,\pi^3}\,
\bigg\{ \bigg[ y^2(3-2y) + {5\lambda_1\over 3m_b^2}\, y^3
  + {\lambda_2\over m_b^2}\, y^2(6+5y) \bigg]\, 2\theta(1-y) \nn\\*
&&\qquad\qquad\quad\ \,\, - \bigg[ {\lambda_1\over 6m_b^2} 
  + {11\lambda_2\over 2m_b^2} \bigg]\, 2\delta(1-y)
  - {\lambda_1\over 6m_b^2}\, 2\delta'(1-y) + \ldots \bigg\} . 
\eeqa
The behavior near $y=1$ is determined by the leading order structure function,
which contains the terms $2 [\theta(1-y) - \lambda_1/(6m_b^2)\, \delta'(1-y) +
\ldots]$.  The derivative of the same combination occurs in the $B\to
X_s\gamma$ photon spectrum~\cite{FLS}
\beqa\label{bsgspec} 
{\d \Gamma\over \d x} &=& 
  {G_F^2\, m_b^5 \,|V_{tb}V_{ts}^*|^2\, \alpha\, C_7^2 \over32\,\pi^4}\, 
  \bigg[ \bigg( 1 + {\lambda_1-9\lambda_2\over2m_b^2} \bigg)\, \delta(1-x) 
  - {\lambda_1 + 3\lambda_2 \over 2m_b^2}\, \delta'(1-x) \nn\\
&&\qquad\qquad\qquad\qquad\quad 
  - {\lambda_1 \over 6m_b^2}\, \delta''(1-x) + \ldots \bigg] .
\eeqa
At subleading order, proportional to $\delta(1-y)$ in Eq.~(\ref{slspec}) and to
$\delta'(1-x)$ in Eq.~(\ref{bsgspec}), the terms involving $\lambda_2$ differ
significantly, with a coefficient $11/2$ in Eq.~(\ref{slspec}) and $3/2$ in
Eq.~(\ref{bsgspec}).  Because of the 11/2 factor, the $\lambda_2\, \delta(1-y)$
term is important in the lepton endpoint
region~\cite{subltwistus,subltwist,subltwistmn}, giving rise to an order 10\%
uncertainty.  There may also be a sizable uncertainty at sub-subleading order
from weak annihilation~\cite{Voloshin,subltwistus}, discussed below.

In contrast to the above, in the $q^2 > (m_B-m_D)^2$ region the first few terms
in the OPE determine the rate with no dependence on the shape
function~\cite{BLL1}.  This is the only differential rate known to order
$\alpha_s^2$~\cite{melnikov}.  The $q^2$ cut implies $E_X \lsim m_D$ and $m_X
\lsim m_D$, and therefore the $m_X^2 \gg E_X \lqcd \gg \lqcd^2$ inequality is
satisfied.  This relies, however, on $m_c \gg \lqcd$, and so the OPE is
effectively an expansion in $\lqcd/m_c$ in this region~\cite{neubertq2}.  The
uncertainties come from order $\lqcd^3/m_{c,b}^3$ nonperturbative corrections,
the $b$ quark mass, and the perturbation series.  Weak annihilation (WA)
suppressed by $\lqcd^3/m_b^3$ is important, because it enters the rate as
$\delta(q^2-m_b^2)$~\cite{Voloshin}.  Its magnitude is hard to estimate,
because it is proportional to the difference of two matrix elements, which are
equal in the factorization limit.  Assuming a 10\% violation of factorization,
WA could be $\sim$\,2\% of the $B\to X_u\ell\bar\nu$ rate, and in turn
$\sim$\,10\% of the rate in the $q^2 > (m_B-m_D)^2$ region.  The uncertainty of
this estimate is large.  Since this contribution is also proportional to
$\delta(E_\ell - m_b/2)$, it is even more important in the lepton endpoint
region.  Experimentally, WA can be constrained by comparing $|V_{ub}|$ measured
from $B^0$ and $B^\pm$ decays, and by comparing the $D^0$ and $D_s$
semileptonic widths~\cite{Voloshin}.

Combining the $q^2$ and $m_X$ cuts can significantly reduce the theoretical
uncertainties~\cite{BLL2}.  The right plot in Fig.~\ref{fig:dalitz} shows that
the $q^2$ cut can be lowered below $(m_B-m_D)^2$ by imposing an additional cut
on $m_X$.  This raises the scale of the expansion to $m_b\lqcd/(m_b^2-q_{\rm
cut}^2)$, resulting in a significant decrease of the uncertainties from both
the perturbation series and from the nonperturbative corrections.  At the same
time the uncertainty from the $b$ quark light-cone distribution function only
turns on slowly.  The results in Table~\ref{tab:dblcut} show that it may be
possible to determine $|V_{ub}|$ with a theoretical error at the $\sim$\,5\%
level using up to $\sim$\,45\% of the rate.  Such accuracy can also be achieved
with cuts somewhat removed from the $b\to c$ threshold.  The experimental
status of these measurements was reviewed elsewhere at this
conference~\cite{thorndike}.

\begin{table}[h]
\centerline{\small
\begin{tabular}{|ccc|}
\hline
Cuts on $q^2$ and $m_X$~  &  ~Fraction of events~
  &  ~Error of $|V_{ub}|$ [$\sigma(m_b) = 80/30\,\MeV$]\\ \hline
$(m_B-m_D)^2,\, m_D$ 		& $17\%$ & $15\%/12\%$ \\ \hline
$6\,\GeV^2,\, m_D$		& $46\%$ & $8\%/5\%$  \\
$8\,\GeV^2,\, 1.7\,\GeV$ 	& $33\%$ & $9\%/6\%$  \\ \hline
\end{tabular}
}
\caption{$|V_{ub}|$ from combined cuts on $q^2$ and $m_X$~\cite{BLL2}.}
\label{tab:dblcut}
\end{table}

In the future there are several ways to reduce the uncertainties: (i) it is
important to try to get the experimental cuts as close to the charm threshold
as possible; (ii) more precise determinations of $m_b$ will be useful, since
the decay rate is proportional to $m_b^5$, and this sensitivity is even
stronger in the presence of cuts; (iii) constrain weak annihilation by
comparing $|V_{ub}|$ extracted separately from $B^\pm$ and $B^0$ decays, or by
comparing the $D^0$ and $D_s$ semileptonic widths; (iv) improve measurement of
$B\to X_s\gamma$ photon spectrum (lower cut) and use it directly in the
analysis of the $E_\ell$ or $m_X$ spectra instead of via intermediate
parameterizations; (v)~calculate the full $\alpha_s^2$ corrections (beyond
$\alpha_s^2\beta_0$), which is only known for the total rate and the $q^2$
spectrum, but not for others.  Clearly, the different $|\Vub|$ determinations
have different advantages and different sources of uncertainties.  One needs to
measure $|V_{ub}|$ in several ways to gain confidence that the uncertainties
are as small as estimated.

\section{Summary}

\begin{itemize} \vspace*{-6pt}\itemsep -2pt

\item $|V_{cb}|$ is known at the $\sim$\,4\% level, error may soon become half
of this.  The inclusive measurement can be improved with more precise and
consistent data on spectral moments; while the exclusive determination needs
${\cal F}_{(*)}(1)$ from unquenched lattice QCD.

\item Model independent determination of $|V_{ub}|$ with $\sim$\,10\% error
seems possible in the near future.  To improve the inclusive determination,
neutrino reconstruction with large statistics is crucial; the exclusive needs
unquenched lattice form factors or use of double ratios.

\item For both $|V_{cb}|$ and $|V_{ub}|$, it is important to pursue both
inclusive and exclusive measurements, as they provide powerful crosschecks.

\item Progress in SCET in understanding $B\to \pi/\rho\, \ell\bar\nu,\
K^*\gamma,\ K^{(*)}\ell^+\ell^-$ form factor relations in the $q^2 \ll m_B^2$
region and its experimental tests will affect the sensitivity to new physics in
these decays, and may also impact our understanding of charmless nonleptonic
decays.

\item The theoretical limit in determining $|V_{cb}|$ and $|V_{ub}|$ (without
lattice QCD) appear to be about 1\% and 4\%, respectively (achieving these
might require a super-B-factory).

\end{itemize}\vspace*{-8pt}

\subsubsection*{Acknowledgments}

It is a pleaseure to thank Christian Bauer, Ben Grinstein, Adam Leibovich,
Mike Luke, Aneesh Manohar, and Mark Wise for many enjoyable collaborations on
the topics discussed.  Discussions with Iain Stewart, Urs Langenegger and
Oliver Buchmuller are also greatly appreciated.
I would like to thank Luis Oliver and the organizers for the invitation to 
a very enjoyable conference.
This work was supported in part by the Director, Office of Science, Office of
High Energy and Nuclear Physics, Division of High Energy Physics, of the U.S.\
Department of Energy under Contract DE-AC03-76SF00098 and by a DOE Outstanding
Junior Investigator award.

%
\label{LigetiEnd}
 
\end{document}